\documentclass[11pt,preprint,preprintnumbers,amsmath,amssymb,nofootinbib,aps]{revtex4}
\usepackage{graphics,color,array,dcolumn}
\usepackage{hyperref}
\usepackage{calc}

\usepackage{amsmath}
\usepackage{amssymb}
\usepackage{xspace}

\usepackage{graphicx}
\usepackage{amsfonts}

\begin{document}

\title{Asymptotic safety in gravity and sigma models
\footnote{Talk given at International Workshop on Continuum and Lattice Approaches to Quantum Gravity, 
Brighton, United Kingdom, 17-19 Sep 08. To appear in PoS.}}

\author{Roberto Percacci\footnote{\it on leave from SISSA, via Beirut 4, I-34151 Trieste, Italy.
Supported in part by INFN, Sezione di Trieste, Italy}}
\email{rpercacci@perimeterinstitute.ca}
\affiliation{Perimeter Institute for Theoretical Physics, 31 Caroline St. North, Waterloo, Ontario N2J 2Y5,
Canada}

\begin{abstract}
There are deep analogies between Einstein's theory of gravity
and the nonlinear sigma models. It is suggested that these similarities may extend
also to the ultraviolet behaviour, in the sense that both theories
could turn out to be asymptotically safe.
\end{abstract}

\maketitle

\section{The nonlinear sigma model}

A nonlinear sigma model (NLSM) is a theory whose configurations are maps $\varphi$ from
spacetime to some internal manifold $N$. In most applications to particle physics
$N$ is a coset space $G/H$ and the theory is assumed to be globally 
invariant under the action of the group $G$.
In contradistinction to other scalar fields used in condensed matter and
particle physics, such as the Higgs model and its generalizations,
the nonlinear sigma models are nonlinear already at the kinematical level,
in the sense that there is no linear structure on the configuration
space: the sum of two fields is not defined. This has deep consequences.
If we limit ourselves to terms with two derivatives, we can write the (Euclidean) action
\begin{equation}
\label{actionnlsm}
\frac{1}{2g^2}\int d^dx\,\partial_{\mu}\varphi^\alpha\partial^\mu\varphi^\beta 
h_{\alpha\beta}(\varphi)\,,
\end{equation}
where we have chosen a coordinate system on $N$, and $h_{\alpha\beta}$ is a metric on $N$
\footnote{for various ways of describing these models see (\cite{perbook})
or (\cite{ps}).}.
A free field theory is one for which the action is quadratic in the field.
This would require that the components $h_{\alpha\beta}$ are independent of $\varphi$,
{\it i.e.} that the metric is flat. But topology forbids the existence of
a flat metric on most manifolds. Therefore the absence of a linear structure
generally implies that these theories can not be free.

Since $h_{\alpha\beta}$ are in general nonpolynomial functions
of the fields, the fields themselves, as well as $h_{\alpha\beta}$,
have to be dimensionless, and consequently the constant $g^2$
has dimension $\mathrm{length}^{d-2}$.
If $N$ is a coset space and the action is $G$-invariant,
a potential must necessarily be constant and can be set
to zero without loss of generality.
Any term in the Lagrangian for such a theory must contain derivatives of the fields,
and (\ref{actionnlsm}) is the most general one containing two derivatives. 
The ground states of the NLSM correspond to constant fields,
and in the absence of a potential they are completely degenerate. 
We will choose an arbitrary vacuum with coordinates $\varphi^\alpha_0$.
Note that a choice of vacuum breaks the global symmetry group $G$ leaving
a residual symmetry $H$. The fields $\varphi^\alpha$ are
the resulting Goldstone bosons, and the NLSM is a theory of Goldstone bosons.

In general, the metric $h_{\alpha\beta}$ can be seen as representing 
infinitely many coupling constants.
To make this manifest, suppose we apply perturbation theory to the NLSM.
First, in order to give the scalar fields their canonical dimension
we absorb the constant $g^2$ in the fields, defining 
$\bar{\varphi}^{\alpha}\!=\varphi^{\alpha}/g$. 
The dimension of $\bar\varphi$ is then
$\mathrm{length}^{2-d\over 2}$. 
Now the action reads
\begin{equation}
\label{redef}
{1\over 2}\int d^dx\,\partial_{\mu}\bar{\varphi}^{\alpha}
\partial^{\mu}\bar{\varphi}^{\beta}
h_{\alpha\beta}({g\bar{\varphi}})\ .
\end{equation}
In order to separate the kinetic term from the
interaction terms we expand the field around the vacuum:
$\bar{\varphi}^{\alpha}\!=\!\bar{\varphi}^{\alpha}_0\!+\!\eta^{\alpha}$,
and expand the metric in Taylor series in $\eta$:
\begin{equation}
\label{expansion}
h_{\alpha\beta}(g{\bar{\varphi}})\!=\!
h_{\alpha\beta}(g{\bar{\varphi}_0})\!+\!
g\,\partial_{\gamma}h_{\alpha\beta}(g{\bar{\varphi}_0})\eta^{\gamma}
+\frac{1}{2}
g^2\partial_{\gamma}\partial_{\delta}h_{\alpha\beta}(g{\bar{\varphi}_0})
\eta^{\gamma}\eta^{\delta}+\cdots
\end{equation}
where we write $\partial_\gamma$ for ${\partial\over\partial\varphi^\gamma}$.
The coefficients of this expansion are
now field-independent and represent the coupling constants of the theory.
Note that there is in general an infinite number of couplings and all
couplings involve derivatives of the fields. The dimension of the coupling
constant in the $m$-th interaction term, i.e. the coefficient of
$\partial\eta\,\partial\eta\,\eta^m$, is
$\mathrm{length}^{{m\over 2}(d-2)}$. 
In spite of the infinite number of couplings, this theory is renormalizable 
in a generalized sense for $d\!=\!2$ \cite{friedan}. 
It is power counting nonrenormalizable for $d\!>\!2$.
Therefore, the nonlinear sigma model is not a good candidate for a
fundamental theory in more than two dimensions. 

The existence of symmetries greatly reduces the number of couplings.
The action (\ref{actionnlsm}) is $G$-invariant if the metric $h_{\alpha\beta}$
is $G$-invariant, which means that there exist Killing vectors $K_a$
satisfying the Lie algebra of $G$.
For example, if $N$ is topologically a semisimple Lie group $G$, 
and the metric $h_{\alpha\beta}$ is invariant under both left and right multiplications,
then $h_{\alpha\beta}$ is unique up to an overall scale.
In this case $G$-invariance reduces the number of arbitrary coupling constants
to just one, which can be identified with $g$.
In this case it is common to describe the NLSM in a different formalism:
choose a linear representation of $G$, and a basis $\{T_a\}$
in the Lie algebra such that the invariant inner product is 
$\mathrm{Tr}T_aT_b=\delta_{ab}$.
The element of the group with coordinates $\varphi^\alpha(x)$ corresponds
to a matrix $U(x)$. Then the action (\ref{actionnlsm}) can be rewritten in the form
\begin{equation}
\label{action2}
\frac{1}{2g^2}\int d^dx\,\mathrm{Tr}\,(U^{-1}\partial_{\mu}U\,U^{-1}\partial^\mu U)
\end{equation}
This form exhibits most clearly the invariance under global left and right $G$
transformations.
Since the field $U=\mathbf{1}$ is invariant under the diagonal subgroup
$\Delta G=\{(g,g)\}$, the proper characterization of the manifold $N$ in this
case is as the coset $(G_L\times G_R)/\Delta G$,
where $G_L$ and $G_R$ denote the group $G$ acting on itself from left or right
\footnote{Note that there is a subtle
difference between the Lie group $G$ and this coset: the former has a preferred
element, namely the identity, whereas the  latter does not.}.
For this reason, these are often called ``chiral models''.

\section{Chiral perturbation theory}

The main application of the NLSM to particle physics is as a low energy phenomenological 
approximation of QCD. In this case the field $\varphi$ takes values in the
manifold $(SU(2)_L\times SU(2)_R)/\Delta SU(2)$, 
and represents the direction chosen by the quark condensate.
The three (pseudo) scalar fields are identified with the pions, and they
can be thought of as low energy fluctuations in the condensate.
The coupling $g$ is equal to $1/f_\pi$, where $f_\pi$ is the pion decay constant.
This model is an effective low energy description of strong interactions.
A slightly more general, but less accurate, model where $SU(2)$ is replaced by $SU(3)$,
describes the dynamics of the octet of mesons, including also the kaons and the $\eta$.
These models cease to give a good description of meson interactions
at an energy scale of the order of several hundred MeV's.
One can see this breakdown from several viewpoints:
\begin{enumerate}

\item
experimentally, it is the scale at which other strongly interacting particles appear,
which are not accounted for by the model;
\item
from the point of view of QCD it is the characteristic mass scale of the condensate,
so that there is no meaning in talking of the dynamics of the condensate above this scale;
\item
from the point of view of the effective field theory itself,
it is the scale at which perturbation theory breaks  down, as we shall discuss below.
\end{enumerate}

Besides pion physics, there is also another important application of the NLSM.
If we use the local isomorphisms $SU(2)\sim SO(3)$ and $SU(2)\times SU(2)\sim SO(4)$ 
to identify $N\approx SO(4)/SO(3)\approx S^3$, we can identify this manifold 
with the orbit of the minima of the potential of the complex Higgs doublet of the standard model.
This NLSM can be viewed as a low energy effective theory describing the dynamics of the
electroweak Goldstone bosons, in the approximation in which the
mass of the Higgs particle is infinitely large \cite{ab}.
Since the Higgs mass is proportional to the quartic self coupling in
the Higgs potential, this is often referred to as the strong coupling limit
of the Higgs model.
When this NLSM it is coupled to $SU(2)_L\times U(1)$ gauge fields, the Goldstone bosons
give mass to the $W^\pm$ and the $Z$, and in the unitary gauge
nothing is left in the scalar sector. 
Thus, the model can be viewed as describing a ``Higgsless Higgs phenomenon''.
In this case we can identify $g=1/\upsilon$, where $\upsilon=246$\,GeV is the Higgs VEV.
Above this scale the model is expected to break down at least for the third of the reasons 
listed above.
But unlike the case of strong interactions, the corresponding energy scale has not
yet been explored experimentally, so we don't know whether the other
two reasons also hold.

There is a general formalism that can be used to extract physical consequences from
these effective field theories, going under the name of ``chiral perturbation theory''
or $\chi$PT for short \cite{weinberg}.
The NLSM is treated as a quantum field theory with a UV cutoff at scale $\Lambda$.
When one computes loop effects with the action (\ref{actionnlsm}),
divergences appear that are proportional to operators containing more than two derivatives.
In order to absorb such divergences it is necessary to assume that the corresponding
operators are present in the action. Therefore in the effective field theory
one has to allow all terms that are compatible with the symmetries of the system.
Schematically, the action can be organized in a derivative expansion
\begin{equation}
\label{expnlsm}
\int d^4x\,\left[g_2 (U^{-1}\partial U)^2+g_4 (U^{-1}\partial U)^4
+g_6 (U^{-1}\partial U)^6+\ldots\right]\,.
\end{equation}
This expression is schematic because there can be several operators
with the same number of derivatives, but for our present purposes, they all behave
in the same way.
Chiral perturbation theory systematically organizes the contributions of all these
operators.
When one studies a process with a given number of external legs
characterized by some external momenta of order $p\ll \Lambda$,
and one wants to calculate e.g. a cross section within a certain theoretical
accuracy, one need to calculate
only a finite number of contributions, where terms with higher derivatives
appear at the same level as operators with fewer derivatives but in
diagrams with higher loops.
For example to correctly describe the experimentally measured pion scattering 
cross sections, one has to supplement the
action (\ref{actionnlsm}) by the four--derivative terms
$$
\int d^4x\,\left[g_{41} \mathrm{Tr}(U^{-1}\partial U)^4
+g_{42} (\mathrm{Tr}(U^{-1}\partial U)^2)^2
\right]
$$
and these four derivative terms give contributions, at tree level, which are
of the same order as those of (\ref{action2}) at one loop.
The couplings $g_2$, $g_{41}$ and $g_{42}$ have to be taken from experiment,
but three experiments suffice to determine them and one can actually do
more than three experiments with pions, 
in such a way that the theory retains some predictive power.

When one increases the accuracy, or the momentum, more and more contributions
have to be taken into account. Eventually when one wants to describe physics near the cutoff
scale, all operators give comparable contributions.
The coefficients of infinitely many operators would then have to be determined by experiment
and as a consequence this effective field theory description would lose its predictive power.

\section{Effective quantum field theory of Gravity}

The historical role of Einstein's theory is as a classical field theory of gravity,
and in this role it has been extraordinarily successful.
Because all other interactions are described by quantum theories,
we would like to understand also gravity in quantum terms.
This poses a number of difficult challenges, both at technical and conceptual level.
But in view of what was said in the previous sections, we should recognize that
a quantum description of gravity does not necessarily have to be valid up to
infinitely high energies. As a first step we could content ourselves with an
effective field theory description holding up to some cutoff scale \cite{burgess}.
In order to understand this better, let us write the (Euclidean) Hilbert action:
\begin{equation}
\label{ehaction}
-\frac{1}{16\pi G}\int d^dx\sqrt{g}\,g^{\mu\nu}R_{\lambda\mu}{}^\lambda{}_\nu\,.
\end{equation}
We note that, omitting indices, the Riemann tensor is schematically of the form 
$\partial\mit\Gamma+\Gamma\Gamma$
and that $\mit\Gamma$ is of the form $g^{-1}\partial g$, so that, aside from the
metrics appearing explicitly in (\ref{ehaction}), the structure of the Lagrangian
is $(g^{-1}\partial g)^2$, exactly as in (\ref{action2}).
As a matter of fact, one can easily see that the metric $g$ is a Goldstone boson
(it has values in the coset $GL(4)/O(4)$) so the similarity of the Lagrangians
is just a reflection of this deeper similarity at the kinematical level
\footnote{Similar statements hold also in the vierbein formulation 
and in more general formulations (\cite{perbook}).}.

Newton's constant has exactly the
same dimension as the coupling $g^2$ in (\ref{action2}), and therefore, 
at the level of power counting, Einstein's theory has
the same quantum properties as the NLSM.
Of course power counting does not tell us whether the potential divergences
do occur or not: for this one has to do the actual calculations. 
These were done, and the nonrenormalizable divergences were seen to 
appear at one loop in the presence of matter \cite{veltman} 
and at two loops for pure gravity \cite{sagnotti}.
As in the NLSM, quantum loops formed with the Hilbert action will therefore
generate infinitely many higher derivative terms. 
Because of diffeomorphism invariance, the effective action will
have, somewhat schematically, the form
\begin{equation}
\label{expgrav}
\int d^dx\sqrt{g}\,\left[g_0+g_2 R+g_4 R^2+g_6 R^3+\ldots\right]
\end{equation}
where $R^m$ has to be interpreted as a generic local, scalar operator
containing $2m$ derivatives of the  metric (for example a scalar constructed
by contracting the indices of $m$ Riemann tensors), 
and the couplings $g_i$ have the same dimensions 
as the homologous couplings in (\ref{expnlsm}).
The main difference between these expansions is the appearance of the
$g_0$ (cosmological) term, which has no analogue in the NLSM.
We can apply to this theory the methods of $\chi$PT \cite{burgess}.
The leading term will be the tree level contribution of the Hilbert action.
This is just the classical result of Einstein's theory.
The first correction will come from one loop terms calculated with the Hilbert action,
cut off at the Planck scale, or from the four derivative terms. 
At momentum scale $p\ll 1/\sqrt{G}$ these corrections will be suppressed
by powers of $p \sqrt{G}$ and since the Planck scale is so much greater than any accessible energy,
these corrections are always extremely small, actually much smaller than the
homologous corrections in pion theory.
From this point of view gravity could be regarded as the best available
example of an effective quantum field theory:
every experimental success of Einstein's classical theory is 
automatically also a success for this effective quantum field theory of 
gravity.
We can therefore say that all we know about gravity is consistent 
with this picture.
But there is also a downside, and that is that all we know about gravity is also
consistent with treating it just as a classical field theory.
At present there seems to be only one possible scenario when quantum corrections 
could become directly testable at high energy accelerators, and this is if there 
exist ``large extra dimensions'',
which would bring the effective Planck scale down to the TeV \cite{large}.

\section{Asymptotic safety}

By definition, effective field theories break down near 
their UV cutoff and have to be replaced by another theory at higher energies.
The paradigm of this transition is again the theory of strong interactions.
At high energies they are described by QCD, which is a weakly interacting theory.
Around the GeV scale the QCD interactions become strong, bound states form
and at sufficiently low energies their dynamics is described by the NLSM.
The transition is not fully understood in detail, but few doubt that
this picture is basically correct.
Clearly in this particular application there is not much reason to try
and extend the validity of the NLSM description beyond its natural cutoff.
But in sections 2 and 3 I also mentioned other two low energy effective field theories,
for which the relevant energy scale has not yet been explored.
In the electroweak case we do not yet know for sure that new states 
(whether fundamental particles or composites or resonances)
will be found at colliders,
so we cannot (yet) invoke the first two reasons given in section 2 to claim that the
electroweak chiral model must break down at a scale of a few hundred GeV's.
Even less can be said in the case of gravity.
So in these cases it seems worth exploring the possibility that 
also the third of the given reasons, which is the only one we can use
from our low energy perspective, can be circumvented.
Namely, one may hope that as one approaches the putative cutoff $\Lambda$, 
the effective field theory somehow manages
to heal itself of its perturbative problems and could continue to make sense
also at higher scales, potentially up to infinite energy.
Specifically this can happen if the running couplings, 
expressed in suitable units, reach a Fixed Point (FP).

Let us illustrate this in the case of the NLSM action (\ref{actionnlsm}).
As one sees from the expansion (\ref{expansion}), the effective $n$-point
vertex is proportional to $\tilde g^{n-2}$,
where $\tilde g=g p^{\frac{d-2}{2}}$ is an effective dimensionless coupling,
which takes into account the classical momentum dependence of the interaction.
In perturbation theory % the dimensionful coupling $g$ is independent of $p$, so that 
$\tilde g$ is essentially proportional to $p^{\frac{d-2}{2}}$,
and if one tries to use this theory beyond the cutoff scale,
$\tilde g$ becomes arbitrarily large.
The way in which the theory could cure itself of this problem is if the full 
renormalized $g$, including quantum corrections, 
depended on $p$ in such a way as to exactly compensate this ``classical'' $p$-dependence.
More precisely, we say that the NLSM with action (\ref{actionnlsm}) has a FP if 
$g(p)\sim p^{\frac{2-d}{2}}$
in such a way that $\tilde g(p)$ tends to a constant $\tilde g_*$.

More generally, assume that at a given scale $k$, physics is described by
the tree level expansion of a kind of ``Wilsonian effective action'' 
$\Gamma_k$ which already contains the effect of loops,
the integration over the loop momenta having been extended down 
to the scale $k$. (For this reason we will sometimes refer to the scale $k$ as an ``IR cutoff''.)
We assume that the $\Gamma_k$ has the most general form allowed by the symmetries of the theory 
and that it can be parametrized as in (\ref{expnlsm}) (or (\ref{expgrav}) in the case of gravity),
where $g_i(k)$ are essential renormalized couplings.
The word ``essential'' here refers to the fact that any couplings that can be
eliminated by means of field redefinitions, such as the wave function
renormalization constants $Z$, do not affect physical observables
and therefore need not be considered for our argument.  
We define $\tilde g_i=g_i(k) k^{-d_i}$, where $d_i$ is the canonical dimension of $g_i$.
These dimensionless numbers are the essential couplings measured in units of $k$.
We say that the theory has a fixed point if all the $\tilde g_i$ tend to finite limits
$\tilde g_{i*}$ when $k\to\infty$.

If such a FP exists, and the physical world is described by a RG trajectory 
that hits the FP when $k\to\infty$,
then all the renormalized couplings remain finite when measured in units of $k$,
and the theory has a sensible UV limit.
The important question then is: how many such trajectories exist?
This question is important because if it turned out, for example,
that all trajectories end at the FP, then we would be in the same
situation as with a nonrenormalizable theory:
in order to determine what trajectory the real world corresponds to,
we would have to measure separately each of the infinitely many
couplings of the theory, and the theory itself would have 
little predictive power, if any at all.
At the other extreme, the best possible case would be if there existed a single
such trajectory, because then the demand of having a good UV limit would
constrain all the couplings, up to one parameter telling us the energy scale
corresponding to a particular point on the trajectory.
In order to describe the general case, define the UV critical surface to be the 
set of points in parameter space that is attracted to the FP in the UV limit.
For the field theory description to be valid up to arbitrarily high energy,
the real world must correspond to a point in this surface. 
If the UV critical surface is finite dimensional, then finitely many experiments
are sufficient to determine our position in it. 
Then, the theory is completely pinned down and we can use it to
make predictions.

To summarize: a theory that has a FP with finitely many attractive directions
has a sensible UV limit and is predictive.
Such a theory is called ``asymptotically safe'' \cite{Weinberg}.
QCD is the prime example; in this case the FP corresponds to a free theory
and one has the added bonus that perturbation theory becomes better and better
as energy increases.
But it is clear that an asymptotically safe theory would be perfectly sensible even
if the FP was not free.
In this case perturbation theory may be of limited use (if the FP is not too
far away from the free theory FP) or of no use (if the FP corresponds to a
strongly interacting theory).

\section{Simple calculations}

The nonlinear sigma models and gravity have very similar perturbative behaviour.
Do these theories have a FP with the properties that are needed for 
asymptotic safety in $d=4$?
Various approximation schemes have been applied, and together they
provide support for this hypothesis.
One approach that has been pursued is to truncate the effective action $\Gamma_k$,
i.e. keeping only a finite number of terms in the expansions (\ref{expnlsm})
or (\ref{expgrav}), and to gradually increase the number of term.
For historical reasons, more work has been done on gravity than on the NLSM,
so that the evidence for asymptotic safety is now stronger for the former than for the latter.
Here I will mention only the evidence that is presently available
for both theories, which amounts to truncating the effective action
to the lowest terms, those at most quadratic in derivatives.
I refer to \cite{reviews} for more detailed reviews and references.

As mentioned in the preceding section, we must first make sure that the
action is parametrized just by essential couplings, i.e. that one cannot
eliminate them by field redefinitions.
Here the difference between the two theories manifests itself.
In the case of the NLSM, if we try to absorb the coupling $1/g^2$ by a
redefinition of the field, as in (\ref{redef}), then $g$ reappears in
the internal metric $h_{\alpha\beta}$, as in the expansion (\ref{expansion}).
Thus $g$ is clearly an essential coupling.
On the other hand, as we have already mentioned, in the case of the
Hilbert action (\ref{ehaction}) the field (in this case the metric)
also appears in the volume element and in the contraction of indices.
As a consequence the effect of rescaling the field is quite different
and Newton's constant can be completely absorbed
by the redefinition $g_{\mu\nu}\to 16\pi G g_{\mu\nu}$
(the Ricci tensor is not affected by this redefinition).
The reason why $G$ can still be considered an essential coupling is that,
uniquely among field theories,
a rescaling of the metric also affects the definition of the cutoff
and if we decide to use the cutoff as the unit of energy,
as was implicit in our preceding discussion,
then such a rescaling is not allowed \cite{perini3}.
So it is true also in gravity, when parametrized by the Hilbert action
and using cutoff units, 
that Newton's constant is an essential coupling.

The tool that has been used in the last ten years or so to calculate 
the beta functions of these nonlinear theories is a form of 
Exact Functional Renormalization Group Equation
(FRGE) introduced by Wetterich \cite{Wetterich}. 
This is a very powerful and versatile tool, which by now has been applied
to a large variety of problems \cite{FRGE}, but since it is not yet very widely known,
there may be the impression that in order to see the appearance of a FP
in gravity one has to resort to somewhat esoteric methods.
But this is not so: one can see the FP already in the lowest order of perturbation theory. 
Both in gravity and in the NLSM, restricting ourselves to the terms with two derivatives,
the coefficient $g_2$ that multiplies the action has dimension of $\mathrm{mass}^{d-2}$.
Therefore the leading term in the beta function is expected to be of the form
\begin{equation}
\label{betaoneloop}
k\frac{\partial g_2}{\partial k}=B_1 k^{d-2}\ ,
\end{equation}
where $B_1$ is a calculable numerical coefficient.
(In a direct calculation of the effective action, this beta function would show up
as a power law divergence.)
The (square of the) perturbative coupling constant is proportional to the inverse 
of $g_2$: $g^2=1/2g_2$.
Therefore its beta function is of the form
$$
k\frac{\partial g^2}{\partial k}=-2B_1g^4 k^{d-2}
$$
and the beta function of the dimensionless coupling $\tilde g^2=g^2 k^{d-2}$ is
\begin{equation}
\label{dimlessbetaoneloop}
k\frac{\partial \tilde g^2}{\partial k}=(d-2)\tilde g^2-2B_1\tilde g^4\ .
\end{equation}
A nontrivial, UV attractive fixed point will exist provided $B_1>0$.
This then is the basic mechanism for the appearance of a FP.
Of course, since the FP may correspond to a relatively large value of the coupling,
one cannot in general give too much weight to this perturbative result.
It is for this reason that one has to resort to a more powerful method such as the FRGE.
In practice, the actual calculation based on the FRGE is somewhat similar to the one loop 
calculation,
but the approximation upon which it is based does not rely on the coupling being small.
This makes it an ideal tool to explore the notion of asymptotic safety.

There is a way of approximating the truncated FRGE that will reproduce exactly
the one loop result: it consists in neglecting the running of the couplings
in the r.h.s. of the equation.
When this is applied to the NLSM or to gravity, one gets (\ref{dimlessbetaoneloop}).
The truncated FRGE without this one loop approximation yields a result of the form:
\begin{equation}
\label{dimlessbeta}
\frac{d\tilde g^2}{dt}=(d-2)\tilde g^2-\frac{2B_1\tilde g^4}{1-B_2\tilde g^2}\ ,
\end{equation}
where $B_2$ is another calculable numerical coefficient.
Further improvements in the applications of the FRGE consist not in adding
higher loops formed with the two derivative operator, but in adding the
contributions of higher derivative operators.
One should not expect these additional contributions to be numerically small.
However one may hope that at least the results for the FP-values and the critical
exponents of the lower order operators are not dramatically shifted by the addition
of the new operators.
For some evidence that in gravity this is indeed the case, I refer to the
talk by C. Rahmede.

\subsection{Nonlinear sigma model}

Let me now give some explicit results, starting from the NLSM. 
The application of the FRGE makes use of background field techniques
developed in \cite{honerkamp}. 
Here we follow \cite{codello2}.
Having chosen a (not necessarily constant) background $\bar\varphi$,
for each $x$ one can find a unique vector $\xi(x)$ tangent to $\bar\varphi(x)$
such that $\varphi(x)$ is the point on the geodesic passing through
$\bar\varphi(x)$ and tangent to $\xi(x)$, the distance from $\bar\varphi(x)$
being equal to $|\xi(x)|$.
One should think of $\xi$ as an element of the space of smooth sections
$C^\infty(\bar\varphi^*TN)$,
and we write $\varphi(x)=Exp_{\bar\varphi(x)}\xi(x)$, where $Exp$ is the exponential map.
The action for $\varphi$ can be rewritten as $S(\varphi)=\bar S(\bar\varphi,\xi)$.
The second variation of the action yields an operator $\Delta$ acting on $\xi$ by
$\Delta(\xi)^\alpha=-\bar D^2\xi^\alpha
-\partial_\mu\bar\varphi^\gamma\partial^\mu\bar\varphi^\delta 
\bar R^\alpha{}_{\gamma\beta\delta}\xi^\beta$,
where 
$\bar D_\mu\xi^\alpha=
\partial_\mu\xi^\alpha+\partial_\mu\bar\varphi^\gamma\bar{\mit\Gamma}_\gamma{}^\alpha{}_\beta\xi^\beta$ 
is the covariant derivative constructed with the Christoffel symbols 
of $h_{\alpha\beta}$, and $\bar R_{\alpha\gamma\beta\delta}$ its Riemann tensor. 
The bar indicates that everything is to be evaluated at the background field.
One can check explicitly using the transformation properties 
\begin{equation}
\xi'^\alpha={\partial\bar\varphi^{\prime\alpha}\over\partial\bar\varphi^\beta}\xi^\beta\ ;\qquad
\bar{\mit\Gamma}'{}_\lambda{}^\mu{}_\nu%({\bar\varphi}^\prime)
={\partial \bar\varphi^\tau\over\partial \bar\varphi^{\prime\lambda}}
{\partial \bar\varphi^{\prime\mu}\over\partial \bar\varphi^\rho}
{\partial \bar\varphi^\sigma\over\partial \bar\varphi^{\prime\nu}}
\bar{\mit\Gamma}_\tau{}^\rho{}_\sigma%(\bar\varphi)
+{\partial \bar\varphi^{\prime\mu}\over\partial \bar\varphi^\rho}
{\partial^2 \bar\varphi^\rho\over\partial \bar\varphi^{\prime\lambda}\partial \bar\varphi^{\prime\nu}}
\end{equation}
that the covariant derivative transforms in the same way as $\xi$ under
diffeomorphisms of $N$.
One defines an IR cutoff by adding to the bare action a nonlocal term of the form
\begin{equation}
\label{cutoff}
\Delta S_k(\bar\varphi,\xi)=
\frac{1}{2g^2}\int d^dx\,
h_{\alpha\beta}\xi^\alpha R_k(-\bar D^2)\xi^\beta\ ,
\end{equation}
for a suitable profile function $R_k(z)$ which goes to zero sufficiently fast for $z>k$,
and to $k$ for $z\to 0$.
The beta function of ${\tiny\frac{1}{g^2}}h_{\alpha\beta}$ is then given by the FRGE as a trace of a certain function of the operator $-\bar D^2$.
At one loop one finds in general that it is proportional to the Ricci tensor \cite{codello2}.

\begin{figure}
\includegraphics[width=.6\textwidth]{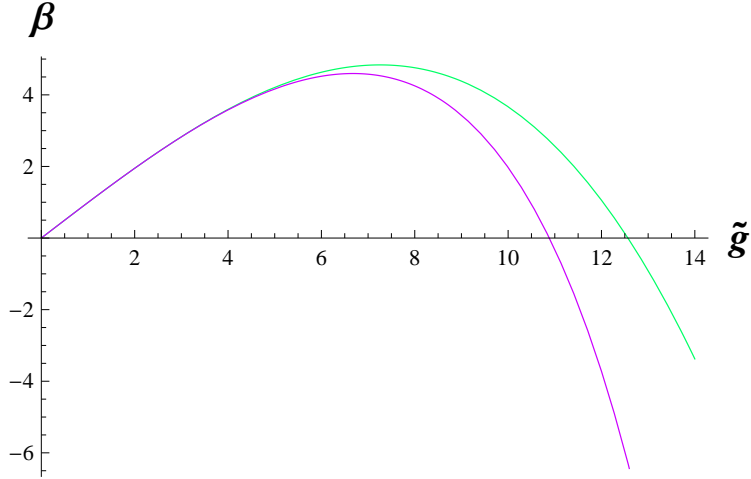}
\caption{The one loop beta function (upper curve) and the beta functions
extracted from the truncated FRGE (lower curve) for the case $N=S^3$,
which corresponds to the chiral model of pions or electroweak bosons.
Chiral perturbation theory applies near the origin.}
\label{fig1}
\end{figure}

If there are any symmetries, the flow will preserve them.
To see this, let $g$ be an element of the symmetry group 
$G$ and $L_g$ the diffeomorphism of $N$ corresponding to $g$.
Since $h_{\alpha\beta}$ is $G$-invariant, also its connection is,
so $L_g$ maps the geodesic through $y$ tangent to $\xi$
to the geodesic through $L_g(y)$ tangent to $TL_g(\xi)$ \cite{kn}:
$L_g(Exp_y(\xi))=Exp_{L_g(y)}(TL_g(\xi))$.
We call $\varphi'=L_g\circ\varphi$ the transform of $\varphi$ under $g$
and $\xi'=TL_g(\xi)$ the transform of $\xi$ under $g$.
Then 
$\varphi'=L_g(Exp_{\bar\varphi}\xi)=Exp_{L_g(\bar\varphi)}(TL_g\xi)
=Exp_{\bar\varphi'}\xi'$.
There follows that
\begin{equation}
\bar S(\bar\varphi',\xi')=S(\varphi')=S(\varphi)=\bar S(\bar\varphi,\xi)\,,
\end{equation}
{\it i.e.} the background field action $\bar S$ is $G$-invariant provided both
background and quantum field are transformed.
%Now let $\Delta^\prime$ be the operator acting on 
%$\xi'\in C^\infty(\bar\varphi^{\prime*}TN)
%=C^\infty(\bar\varphi^*L_g^*TN)$, constructed with $\bar\varphi'$.
Furthermore, we have $-\bar D^{\prime2}(\xi')=TL_g(-\bar D^2(\xi))$
or abstractly $-\bar D^{\prime2}=TL_g\circ(-\bar D^2)\circ TL_g^{-1}$.
There follows that if $\xi$ is an eigenvector of $-\bar D^2$ with eigenvalue $\lambda$,
then $\xi'$ is an eigenvector of $-\bar D^{\prime2}$ with the same eigenvalue.
Therefore the spectrum of $-\bar D^2$ is $G$ invariant.
The FRGE gives the scale variation of $\Gamma_k(\bar\varphi)$ as a sum
over eigenvalues of $-\bar D^2$,
so it follows from the argument above that $\partial_t\Gamma_k(\bar\varphi)$
is $G$-invariant.
This implies that if the starting action $\Gamma_{k_0}(\bar\varphi)$ is $G$-invariant,
also the action at any other $k$ is.

Now assume that $N$ is a $D$-dimensional maximally symmetric space,
which implies that
$R_{\alpha\gamma\beta\delta}=
\frac{R}{D}(h_{\alpha\beta}h_{\gamma\delta}-h_{\alpha\gamma}h_{\beta\delta})$,
where $R$ is the curvature scalar.
Modulo trivial redefinitions, we can assume that
$h_{\alpha\beta}$ is normalized so that $|R|=D(D-1)$
\footnote{if $R>0$ this corresponds to a sphere of unit radius; the case $R=0$ is obviously of no interest.} and the only running coupling is $g$.
One then finds
\begin{equation}
\label{nlsmbeta}
B_1=c_d\frac{R}{D}\ ;\quad 
B_2=2c_d\frac{R}{D(d+2)}\ ;\quad 
c_d=\frac{1}{(4\pi)^{d/2}\Gamma\left(\frac{d}{2}+1\right)}\ ,
\end{equation}
A FP will exist if $R>0$.
If we neglect $B_2$, this reproduces old one loop results of 
\cite{friedan,oldnlsm} for $d=2+\epsilon$.
If $d>2$ there is a nontrivial FP at $\tilde g_*^2=\frac{d-2}{2c_d (D-1)}$.
For large $R$ (which in the case of the sphere means large $D$)
it occurs at small coupling, where perturbation theory is reliable.
The one loop critical exponent is $-\frac{d\beta}{d\tilde g}\Big|_*=d-2$.
Since $B_2>0$, its effect is to shift the FP to smaller value
$\tilde g_*^2=\frac{(d^2-4)}{2c_d d (D-1)}$
and to produce a negative pole in the beta function at $g^2=-1/B_2$.
This pole should be of no consequence, since we imagine the theory
to evolve from the nontrivial FP in the UV to the free FP in the IR
(where $\chi$PT applies).
The other effect of $B_2$ is to make the beta function at the FP steeper, 
changing the critical exponent to
$-\frac{d\beta}{d\tilde g}\Big|_*=\frac{2d(d-2)}{d+2}$.
In $d=4$, this gives the mass critical exponent $\nu=3/8$.
Figure \ref{fig1} gives the beta function of the $S^3$--model in $d=4$.

\subsection{Gravity}

Let us now come to gravity in $d$ dimensions \cite{gravfp}.
In this case one has to take into account the complications that follow
from the presence of a gauge group.
We assume that the theory has been gauge fixed in the de Donder gauge,
with gauge parameter $\alpha=1$.
Also, one has to make some specific choice for the way in which the cutoff
is imposed.  We choose a cutoff of type Ia,  in the language of \cite{cpr2}.
The beta function gets contributions from gravitons and from the ghosts.
It is given by equations (\ref{dimlessbeta}), with 
$$
B_1=\frac{4(d^3-15 d^2+12 d-48)}{3(4\pi)^{d/2}\,d\,{\Gamma}(\frac{d}{2})}\ ;\qquad
B_2=\frac{4(d^2-9d+14)}{3(4\pi)^{d/2}(d+2)\Gamma(\frac{d}{2})}\ .
$$
and the identification, $g^2=-8\pi G$.
Due to this minus sign, the FP now exists for $B_1<0$,
which is indeed the case for all $d$.
In particular at one loop for $d=4$ we find $B_1=-11/3\pi$.
It is interesting that this FRGE--based calculation
agrees qualitatively with a one loop calculation in \cite{bjerrum}
who give the following formula for the scale-dependence of Newton's constant:
$$
G(r)=G_0\left[1-\frac{167}{30\pi}\frac{G_0}{r^2}\right]\ ,
$$
Here $r$ is the distance between two gravitating point particles.
If we identify $k=1/ar$, with $a$ a constant of order one,
this would correspond to a beta function
\begin{equation}
\label{betabjerrum}
\beta_{\tilde G}=2\tilde G-a^2\frac{167}{15\pi}\tilde G^2\ .
\end{equation}
The difference in the numerical coefficient highlights that the position 
of the FP is not universal.
Rather, in these calculations, it depends on details of how the
cutoff is imposed, which are akin to renormalization scheme dependence
in perturbation theory.

Due to the minus sign in the relation between $g^2$ and $G$,
and given that $B_2<0$ for $2<d\leq6$, the effect of the correction term 
in the denominator is again to produce a negative pole in the beta function 
at finite $\tilde G>\tilde G_*$, but the effect on the FP is rather modest,
as seen in figure 2.

\begin{figure}
\includegraphics[width=.6\textwidth]{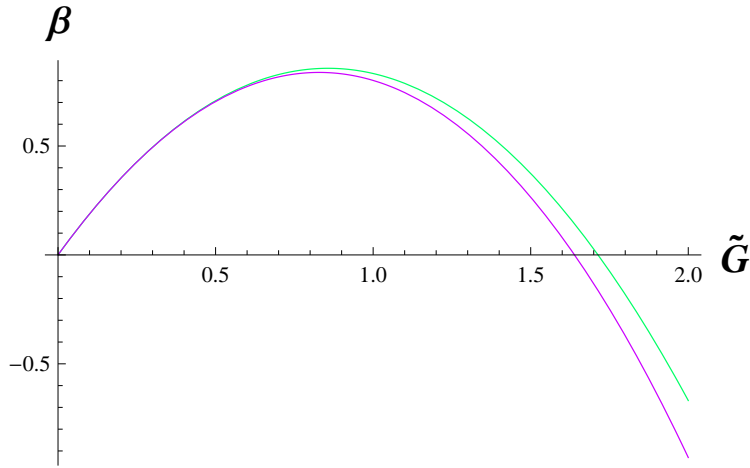}
\caption{The gravitational one loop beta function (upper curve) and the beta function
extracted from the truncated FRGE (lower curve).}
\label{fig2}
\end{figure}

\subsection{Other truncations}

In order to emphasize the similarity with the NLSM,
we have discussed here only the Hilbert action.
But gravity differs from the NLSM in that there can be a term in the action
that does not contain derivatives of the field, namely the cosmological term.
From the point of view of renormalization theory this term is ``more relevant''
than the Hilbert term.
In the presence of the cosmological constant, $B_1$ and $B_2$ become
polynomials in $\tilde\Lambda=\Lambda/k^2$.
The vacuum energy density $\Lambda/(8\pi G)$ depends quartically on $k$, 
and the beta function of $\tilde\Lambda$ itself has the general form
$$
\frac{d\tilde\Lambda}{dt}=
-2\tilde\Lambda+\tilde G\,\frac{A_1+ 2B_1\tilde\Lambda+\tilde G(A_1
B_2-A_2B_1)}{2(1+B_2\tilde
G)}
$$
where $A_1$ and $A_2$ are also polynomials in $\tilde\Lambda$.
The coupled system of beta functions has a FP at positive values
of $\tilde G$ and $\tilde\Lambda$, and since the critical exponents
form a complex conjugate pair, the trajectories approach the
FP with a spiralling motion.
I refer to \cite{cpr2} for more details.

The similarity between the NLSM and gravity extends
also to the case when four derivative terms are taken into account.
In $d=4$ at one loop, in both theories the coefficients $g_4$ of 
equations (\ref{expnlsm}) and (\ref{expgrav}) grow logarithmically, 
so that their inverses, which are the couplings in a perturbative axpansion,
are asymptotically free.
This was shown in a series of papers for gravity \cite{barvinsky} and in \cite{ketov} for the NLSM.
However, in these papers essential use is made of dimensional regularization,
in such a way that information about power-law divergences,
and the corresponding running couplings, is lost.
In the case of gravity, the behaviour of the relevant couplings 
(the cosmological constant and Newton's constant)
in these one loop renormalizable and asymptotically free theories has been
analyzed using the FRGE at one loop in \cite{cp} and it was found that they tend to
a nontrivial FP that is not too far from the one that one finds in the
Einstein--Hilbert truncation.
More recently it has been shown that when one uses the FRGE beyond the one loop approximation,
also the couplings $g_4$ tend to finite limits, so that in a four derivative truncation 
gravity is asymptotically safe but not asymptotically free,
contrary to the one loop results \cite{benedetti}.
Similar calculations for the NLSM are in progress.
In the case of gravity it has also been possible to study higher truncations,
provided one restricts one's attention to powers of the curvature scalar $R$
\cite{lr2,cpr2}.
In the case of the NLSM, this would be analogous to studying actions that
contain only powers of $\mathrm{Tr}(U^{-1}dU)^2$.
In a somewhat different line,
the gravitational RG flow has also been studied in the approximation where
only the conformal degree of freedom is retained, yielding various interesting
results \cite{conformal}.
Finally we mention that a gravitational FP has been
found also using the $\epsilon$ expansion around $d=2$
\cite{epsilon} and the $1/N$ expansion \cite{largen};
this is again entirely analogous to well known results for the
NLSM \cite{oldnlsm}.

\section{Outlook}

Einstein's theory and the NLSM have much in common.
At low energies they can be treated perturbatively as
effective field theories, and I have argued that 
both gravity and certain NLSMs may be asymptotically safe.
If this was the case, then the range of validity of both
theories could extend beyond their natural cutoff scale, 
in principle up to arbitrarily high energy scales.

In general one would expect that the effective action at the FP
$\Gamma_*$ has a very complicated structure, with infinitely many
nonzero couplings.
Still, the theory could make definite predictions.
For example, suppose the critical surface has dimension $d$ and is not parallel to
any of the axes. We can choose the first $d$ couplings $g_1,\ldots g_d$ as
independent variables, and express the remaining ones as functions of these:
$$
g_k=g_k(g_1,\ldots g_d)\ ;\ \ \ \mathrm{for}\ k=d+1,d+2,\ldots
$$
In practice by solving the linearized flow equations one can determine 
the tangent space to the critical surface at the FP:
\begin{equation}
\label{predictions}
g_k=g_{k*}+\sum_{i=1}^d C_{ki}(g_i-g_{i*})\ ;\ \ \ \mathrm{for}\ k=d+1,d+2,\ldots
\end{equation}
For an example of an explicit calculation of this type in a nine-parameter 
approximation to the gravitational effective action, 
I refer again to the contribution by C. Rahmede.
These equations are predictions for the behaviour of the theory at high energy, 
that in principle could be tested: one would have to determine the first $d$ couplings
by means of $d$ independent measurements, and subsequent experiments can
determine whether, for example, the coupling $g_{d+1}$ satisfies the prediction.

Is it reasonable to expect that at the FP only finitely many directions are UV attractive?
In a local quantum field theory there is only a finite number of relevant operators,
whose corresponding couplings have positive mass dimension.
Assuming that the quantum corrections are finite, at most a finite number of
couplings could switch from relevance to irrelevance or vice--versa.
This expectation is supported by the calculations referred to previously.

Let us now mention possible phenomenological applications of this scenario.
There is not much motivation to try to push the pion NLSM beyond its natural cutoff:
we know that at high energies the pions cannot be treated as fundamental particles.
Within $\chi$PT itself, as one goes towards higher energies,
one would have to take into account the heavier mesons, 
and that would change its character drastically. 
For example, the interpretation of the $\sigma$ meson is as the radial component of
a scalar multiplet, and that turns the nonlinear theory into a linear one.
The situation is much less clear in the electroweak version of $\chi$PT, 
where the pions are replaced by the electroweak Goldstone bosons 
(three of the degrees of freedom of the complex Higgs doublet), 
and the pion decay constant is replaced by the VEV $\upsilon=246$\,GeV.
The existence of these Goldstone degrees of freedom has already been established: 
they correspond to the longitudinal components of the W and Z particles.
The general expectation is that a fourth scalar degree of freedom will be found; 
according to the standard model, it is a fundamental scalar field. 
There is no compelling argument against this picture, 
but in principle there is also the possibility that no new degrees of freedom 
will be found at colliders beyond those that we already know.
If this was the case up to sufficiently high energy scales,
then this could be a signal of an asymptotically safe NLSM.

While there is no shortage of models for physics beyond the TeV scale, 
all containing the standard model in the appropriate limit,
there is no quantum field theory valid at and beyond the Planck scale
which likewise subsumes the standard model plus Einstein's theory of gravity.
There is therefore a much stronger motivation for studying asymptotic safety
in the case gravity.
From this point of view one could still see studies of the NLSM as theoretical
exercises where some of the complications of gravity, {\it e.g.} those related to
the gauge structure, are absent.

What would one observe if gravity was asymptotically safe? 
There have been some studies of the phenomenological signatures of asymptotic
safety in gravity \cite{pheno}.
They have considered mosty the case of large extra dimensions,
where the Planck scale is lowered so much that it might be accessible 
to future collider experiments. 
However the same arguments hold also in the four dimensional case, 
just scaling the Planck mass back to its "normal" value.
%Similar behaviour of the cross sections could also emerge in 
%an asymptotically safe NLSM.
It seems more likely that the effects of asymptotic safety could be detected
in early cosmology or in some extreme astrophysical setting.
For some work alomg these lines I refer to the talk by A. Bonanno.

One advantage of the asymptotic safety scenario is that the FP regime could be
the continuation towards higher energies 
of the low energy theory that we are already familiar with, 
without having to change the field content. Although this is not the only
possibility, there could be a smooth transition from the low energy regime,
where all the dimensionful couplings are constant, and $\chi$PT applies,
to the fixed point regime, where all the dimensionless couplings are constant.
In the case of gravity, this would guarantee, by construction, that the UV theory
near the FP corresponds exactly to gravity at low energy, a fact that
is not at all obvious in other, "top down" approaches to quantum gravity.
In this connection we should also mention that the asymptotic safety scenario
need not exclude other "top down" descriptions of quantum gravity. 
In order to make contact with reality, any such theory must first yield
a low energy effective field theory,
and this field theory must match at low energies the $\chi$PT we know.
So, another use of predictions such as (\ref{predictions}) is to check agreement with 
the top down approaches directly at high energy, near the FP regime.
If some sort of agreement could be established, then we would have
much more confidence in both pictures.
Conversely, we observe that the top down approaches are typically based on a 
choice of "classical" action, which cannot be directly identified with the
FP effective average action $\Gamma_*$.
The task of reconstructing the bare action at some UV cutoff scale $\Lambda$
from the effective average action $\Gamma_\Lambda$ (in the limit $\Lambda\to\infty$)
has been discussed in \cite{manrique}.

\medskip
\noindent {\bf Note added}. The beta functions of a NLSM with four derivative
interactions have been studied at one loop in \cite{zanusso}.

\medskip
\noindent {\bf Acknowledgements}. I would like to thank D. Litim for hospitality at
the Department of Astronomy and Physics of the University of Sussex,

\end{document}